\documentclass{IOS-Book-Article}

\usepackage{mathptmx}
\usepackage{soul}\setuldepth{article}
\usepackage{graphicx}
\def\hb{\hbox to 11.5 cm{}}

\begin{document}

\pagestyle{headings}
\def\thepage{}
\begin{frontmatter}              

\title{Intelligent Attacks on Cyber-Physical Systems and Critical Infrastructures}

\renewcommand\thefootnote{}
\footnotetext{\textcopyright\ Alan Sá, Charles Prado, Mariana Flávio, and Luiz Carmo, 2024. The definitive and edited version of this Article is published in NATO Science for Peace and Security Series - D: Information and Communication Security, Volume 65: Modern Technologies Enabling Innovative Methods for Maritime Monitoring and Strengthening Resilience in Maritime Critical Infrastructures, Editors: Pasquale Daponte, Maciej Klósak, Amine Bendarma, online ISBN 978-1-64368-526-7, pages 332 - 351, 2024, DOI: 10.3233/NICSP240033. These details can be found on https://ebooks.iospress.nl.}
\renewcommand\thefootnote{\arabic{footnote}} 

\markboth{}{\hb}

\author[A]{\fnms{Alan Oliveira de} \snm{Sá}\orcid{0000-0001-6311-9672}%
\thanks{Corresponding Author: Alan Oliveira de Sá, alan@di.fc.ul.pt.}},
\author[B]{\fnms{Charles Bezerra} \snm{Prado}\orcid{0000-0003-3119-9772}},
\author[B]{\fnms{Mariana Luiza} \snm{Flavio}\orcid{0009-0009-9477-7163}}
and
\author[B]{\fnms{Luiz F. Rust da C.} \snm{Carmo}\orcid{0000-0001-6131-7771}}

\runningauthor{B.P. Manager et al.}
\address[A]{LASIGE, Faculdade de Ciências, Universidade de Lisboa}
\address[B]{DIEPI, Teaching and Research division, National Institute of Metrology, Quality and Technology (INMETRO), Rio de Janeiro, Brazil}


\begin{abstract}
This chapter provides an overview of the evolving landscape of attacks in cyber-physical systems (CPS) and critical infrastructures, highlighting the possible use of Artificial Intelligence (AI) algorithms to develop intelligent cyberattacks. It describes various existing methods used to carry out intelligent attacks in Operational Technology (OT) environments and discusses AI-driven tools that automate penetration tests in Information Technology (IT) systems, which could potentially be used as attack tools. The chapter also discusses mitigation strategies to counter these emerging intelligent attacks by hindering the learning process of AI-based attacks and points to future research directions on the matter.
\end{abstract}

\end{frontmatter}
\markboth{\hb}{\hb}

\section{Introduction}

Cyberattacks are becoming more complex, frequent, and damaging. At the same time, the tools used by attackers are becoming more advanced, making it easier for them to carry out not only simple but also complex attacks. There are many examples of technology developments obtained over time that are capable of enhancing different types of conventional cyberattacks, ranging from password-cracking to Distributed Denial of Service (DDoS) offensives, among others.
In password-cracking attacks, for instance, the use of GPUs and massively parallel computations together with tools like Hashcat~\cite{Hashcat2024} is greatly increasing the speed and efficiency of password guessing~\cite{aggarwal2018new,alkhwaja2023password}.
In the scope of DDoS attacks, several tools have been developed \cite {kaur2015characterization,de2023distributed} using not only simple flooding techniques, such as TCP SYN and ICMP floods, but also more complex adaptive and multi-vector methods that use network and application layer attacks to increase impact also avoiding detection.

The attack tools and techniques developed to threat Information Technology (IT) systems also find application against Cyber-Physical Systems (CPS) built on Operational Technology (OT). The literature shows the use of widely available tools to attack Programmable Logic Controllers (PLC) that control physical plants in industry and critical infrastructures. Network mapping tools like Nmap \cite{Nmap} and penetration testing tools like Metasploit \cite{Metasploit} can be used, for instance, to execute packet reply attack that modifies the PLC's operations \cite{ramirez2022plc}. WinDbg \cite{Microsoft_windbg}, a Windows debugger for kernel debugging and examining CPU registers and memories, and Scapy \cite{Scapy}, a Python-based packet manipulation tool, can be used to manipulate industrial communication protocol, identify vulnerabilities, and perform exploits in PLCs \cite{hui2021vulnerability}. 
Note that these examples do not limit the plethora of existing modern attack tools and techniques used in both IT and OT environments, and many of these tools are easily obtainable, allowing even non-experts to carry out attacks.

\begin{figure}[ht!]
    \centering
    \includegraphics[trim=1.4cm 3.5cm .9cm 4.1cm, clip=true, width=1\linewidth]{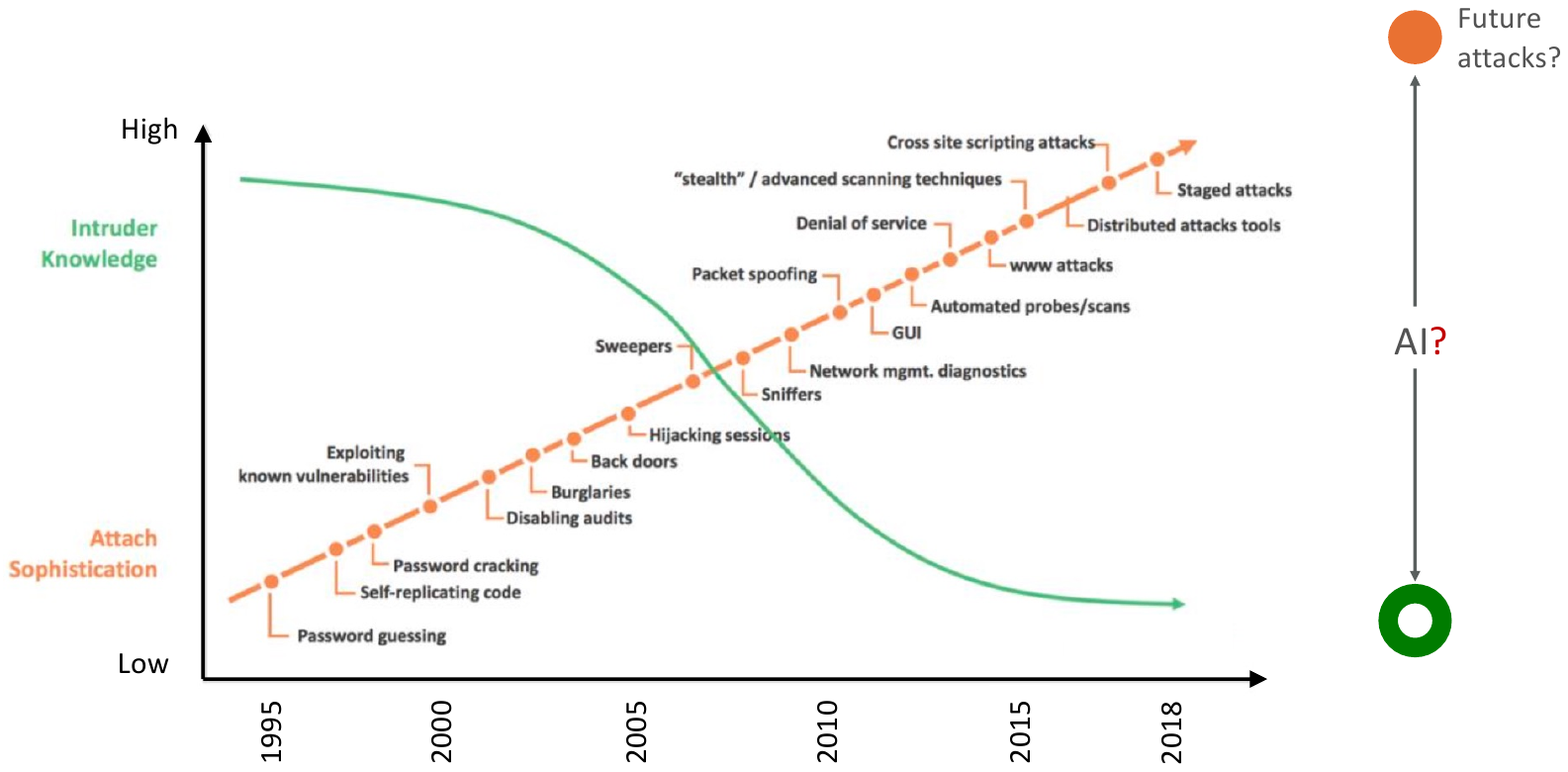}
    \caption{Evolution and trends of cyberattacks. (Adapted from \cite{di2018triton})}
    \label{fig:attack_evolution}
\end{figure}

Figure \ref{fig:attack_evolution} qualitatively depicts two trends that have been observed with regard to the evolution of cyberattacks. It is possible to see that, while the attack sophistication is increasing, the knowledge that attackers need to have when carrying out such attacks has often been reduced. These trends are supported in large part by the attack tools that have been developed over time, as those previously exemplified.

Following in the wake of conventional cyberattacks, recent research leads to a concerning question to be investigated: Can Artificial Intelligence (AI) be used to create intelligent cyber threats, providing attacks with the ability to be stealthy, accurate, and also act on their own? 
The literature points to the development of some AI-based techniques capable of learning information about the target systems and, then, use this knowledge to develop optimized attack actions, hide the attack effects, and assume some of the thinking and decision-making tasks that a human attacker would normally do. This development suggests that AI could potentially handle a great part of the attack process automatically, making sophisticated cyberattacks more commonplace since an attacker, even with limited knowledge, can rely on AI-based tools to multiply its attack fronts.

It is important to note that intelligent attacks, like conventional attacks, can potentially target a wide range of CPS built with common IT and OT. The possible targets can range from industrial plants \cite{humayed2017cyber,mclaughlin2016cybersecurity} to critical infrastructures, such as electric power systems \cite{case2016analysis,sullivan2017cyber}, water canal systems \cite{smith2015covert}, and nuclear plants \cite{mclaughlin2016cybersecurity}, among others. 

In this sense, the objective of this chapter is to provide a detailed review of existing intelligent attack techniques with the potential to affect CPS and critical infrastructures, discuss possible countermeasures, and point to research opportunities on this subject.

The remainder of this chapter is organized as follows: Section \ref{sec:OT} describes existing techniques with the potential to be used in intelligent attacks on OT environments; Section \ref{sec:IT} presents emerging AI-based techniques with potential to automate cyberattacks in IT environments; Section \ref{sec:disc} discusses possible strategies to mitigate intelligent attacks in CPSs; finally, Section \ref{sec:concl} brings the conclusions.

\section{Intelligent Attacks in OT Environments}\label{sec:OT}

As discussed, in CPS and critical infrastructures, the use of OT and their integration into IT has expanded the attack surface to cyber threats, making them prone not only to conventional but also to emerging intelligent cyberattacks. These attacks are characterized by their ability to achieve better offensive capabilities based on acquired knowledge about the target. This greater offensive capability of an intelligent attack can translate into stealth attack actions, increased attack accuracy, or even optimized decisions about attack sequences.

This Section presents a review of the existing intelligent attack techniques in OT environments, detailing their characteristics, aims, and algorithms used. 
To support the discussion about Intelligent attacks in OT environments, Section \ref{subsec:taxonomy} summarizes the existing taxonomy on this subject and explains the roadmap on how such intelligent attacks are built. Understanding this taxonomy and attack roadmap is essential for discussing the possible mitigation strategies to counter these threats. Then, Sections \ref{subsec:CPI_Attacks} and \ref{subsec:MB_Attacks} detail the two main classes of attacks that are commonly integrated to produce an intelligent offensive in OT environments, the Cyber-Physical Intelligence attacks and the Model-based attacks, respectively. 

\subsection{Taxonomy and Attack Roadmap}\label{subsec:taxonomy}

The execution of these attacks in OT systems requires a combination of technical knowledge about networks and the industrial control systems themselves. This includes the ability to navigate through network infrastructure, bypass security measures, and manipulate industrial communication protocols, such as Modbus, Profibus, among others. The complexity substantially increases with the system's level of integration and the sophistication of the defense mechanisms implemented.

According to \cite{de2017covert} the attacks in OT environments can be classified into three main categories: Denial-of-Service (DoS), Service Degradation (SD), and Cyber-Physical Intelligence (CPI) attacks. Note that the concept of service in this taxonomy refers to the physical process executed by the CPS.

The DoS attacks aim to disrupt the service of a system, which can be carried out through the saturation of network and computational resources or arbitrary changes to system data. 
The attack leads to behaviors that might either halt the plant's operations or cause its destruction in the short term \cite{de2017covert}. 
Denial-of-Service (DoS) attacks in OT can have devastating impacts, as by disrupting the service, they can lead to catastrophic failures in critical processes such as water supply systems or power grids. 

The SD attacks, on the other hand, seek to reduce the efficiency or effectiveness of a system, or even reduce its mean time between failures (MTBF), without completely interrupting its operation. 
The attacks are more insidious as their intent is to subtly deteriorate the service, for example, by manipulating the real-time feedback processes, affecting the quality and precision of the control systems \cite{de2017covert}. This can result in the gradual reduction of operational efficiency or increased wear on equipment, leading to premature failures or loss of quality in the service or final product.

Note that both DoS and SD attacks can be implemented by using techniques that cause jitter, packet loss, or false data injection in the CPS network. However, when executed arbitrarily, these attacks can have unpredictable results, ranging from complete ineffectiveness to spurious behaviors beyond those desired by the attacker. 
To guarantee the desired result and prevent unpredictable behavior, the literature indicates the need for these attacks to be built based on accurate knowledge about the models of the target. 
The literature shows that this knowledge is also needed to make the attack stealthy and to decide attack sequences depending on the properties of the target CPS.

The knowledge necessary to develop these types of intelligent attacks is provided through CPI attacks. The CPI attacks include actions that are performed in the automation/control loop of the CPS in order to acquire intelligence about the system, including its physical dynamics, control algorithms, properties, {\it etc}. The CPI attack can be a simple eavesdropping attack (aimed at collecting network traffic), or more sophisticated System Identification (SI) attacks that use algorithms to learn models of the target using eavesdropped data \cite{de2017covert}. The intelligence acquired by CPI attacks can be used to build covert and controlled DoS and SD attacks.

\begin{figure}[ht!]
    \centering
    \includegraphics[trim=4cm 3.1cm 4cm 2.9cm, clip=true, width=1\linewidth]{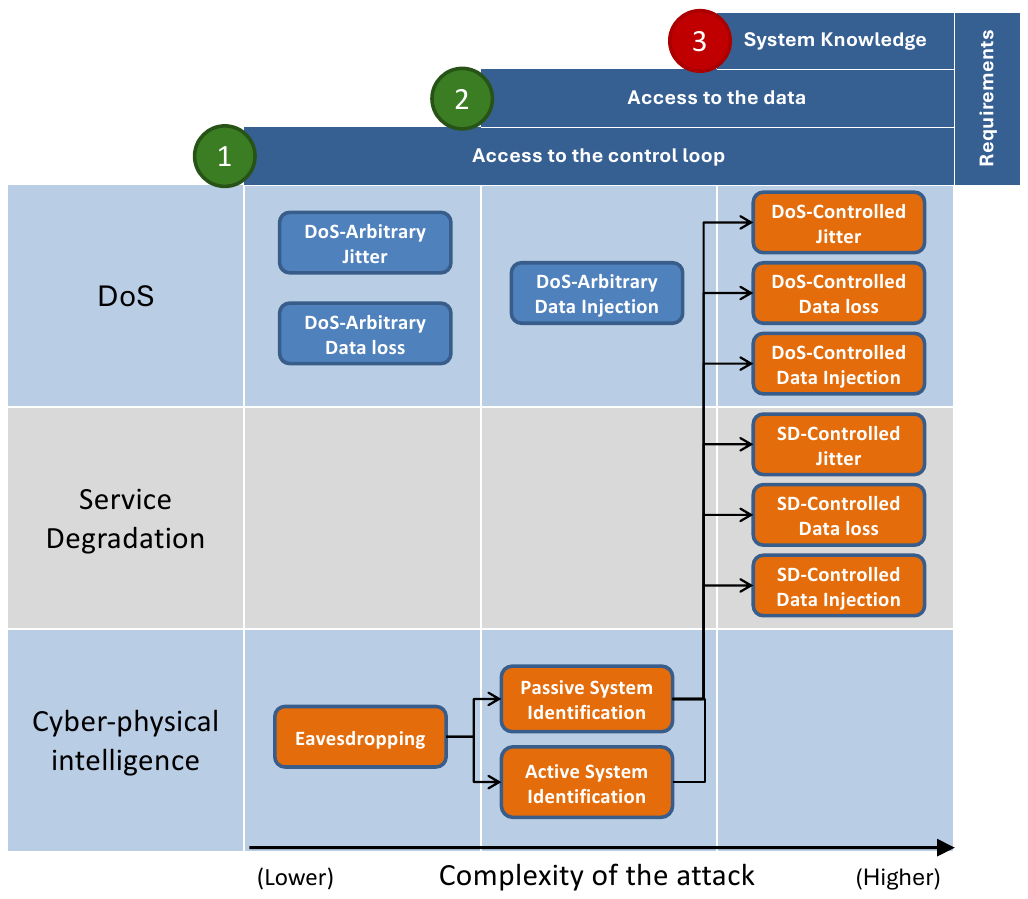}
    \caption{Roadmap for intelligent cyberattacks.}
    \label{fig:attack_roadmap}
\end{figure}

The DoS, SD, and CPI classes of attack are depicted in Figure \ref{fig:attack_roadmap}, presenting the requirements of each attack. 
This framework helps to understand how these different attacks can work together in an orchestrated way to result in an intelligent attack. It also allows us to perceive the level of complexity of the attacks, as well as map what requirements need to be denied to the attacker to protect the system. 
Note that the most complex DoS and SD attacks require knowledge about the targeted system, which is obtained through the CPI attacks. The roadmap to build an intelligent attack in a CPS is highlighted in orange. It is possible to see that the first steps correspond to CPI attacks aimed at satisfying the system knowledge requirement. Then, based on the knowledge acquired by the CPI attack, the intelligent DoS or SD attack is performed.  

In the DoS class, Figure \ref{fig:attack_roadmap} brings six different types of attacks, as conceptualized in \cite{de2017covert}:  DoS-Arbitrary Jitter, DoS-Arbitrary Data Loss, DoS-Arbitrary Data Injection, DoS-Controlled Jitter, DoS-Controlled Data Loss, and DoS-Controlled Data Injection.

The DoS-Arbitrary Jitter attack stands out as a form of DoS attack that disrupts the plant service by creating unpredictable delay variations in data communication. Such disturbances can cause, for instance, the instability of the control process, damage to the physical plant, and the consequent service interruption. As shown in Figure \ref{fig:attack_roadmap}, executing this attack requires access to the control loop.

The DoS-Arbitrary Data Loss attack focuses on causing the loss of crucial data (sensor feedback signals or control commands to actuators) transmitted in the CPS communication links. This attack can be a deliberated network overload leading to packet loss, or more specialized attack strategies that prevent the arrival of important information to the CPS devices. Similar to the DoS-Arbitrary Jitter attack, the DoS-Arbitrary Data Loss attack also requires access to the control loop.

The DoS-Arbitrary Data Injection is more complex than the two previous attacks since it requires access to the data transmitted in the CPS communication network. It means that possible confidentiality protection mechanisms ({\it, e.g.}, symmetric or asymmetric encryption) need to be overcome to implement this attack. Considering that it involves inserting fabricated data into the CPS network, possible data integrity verification mechanisms in place must also be overcome. This action aims to affect the system's behavior, tricking it into performing arbitrarily based on incorrect information. However, since this attack does not rely on any system knowledge, its consequences are hard to predict.

As discussed, the requirements highlighted for executing such attacks are critical for understanding and defending against them. Access to the CPS control/automation loop can be gained if penetration techniques \cite{Valea_2020,Jagamogan_2022,Ghanem_2020,Hu_2020} find exploitable vulnerabilities in the system, allowing the attacker to reach the OT network and directly intervene in the critical communication process of the target. Data access goes further, allowing an attacker not only to intercept but also to read and modify the information in transit. Finally, system knowledge is imperative for executing intelligent attacks such as DoS-Controlled Jitter, DoS-Controlled Data Loss, and DoS-Controlled Data Injection. These attacks require learning details of the CPS, such as its physical models, control algorithms, network properties, {\it etc}. This understanding enables DoS attacks to be precisely calibrated to be effective and cause the intended impact on the target.

In the category of Service Degradation (SD) attacks, similarly to the DoS-Controlled attacks, there are the SD-Controlled Jitter, SD-Controlled Data Loss, and SD-Controlled Data Injection. Note that, as defined in \cite{de2017covert}, the SD-controlled attacks need to be planned based on accurate knowledge of the dynamics of the CPS. Otherwise, the attack can easily evolve into unpredicted behaviors, eventually interrupting the physical process and becoming a DoS attack. Recall that the aim of SD attacks is not to cause an immediate service interruption but to impair the quality of the CPS service and decrease its operational efficiency over time.

The attack known as SD-Controlled Jitter involves meticulous manipulation of data packet transmission times. Unlike an arbitrary jitter attack, which is random and unpredictable, the SD-Controlled Jitter attack is designed to intelligently introduce systematic delays in the CPS communication process, subtly degrading system performance. These delays are calibrated to affect the system without shutting it down, requiring the attack algorithm to have a deep knowledge of CPS models to plan and anticipate the impact of the attack actions.

The SD-Controlled Data Loss attack is characterized by the selected discard of data packets, affecting those that, if lost, cause desired degradation behavior to the CPS. This not only requires a detailed knowledge of system models but also an understanding of how the system responds to the absence of certain data. This allows the attacker to maximize accuracy while keeping attack interventions minimal (aiming stealthiness).

In the SD-Controlled Data Injection attack, the attacker carefully inserts fabricated data within the CPS network to degrade the system service. This strategy is employed to deceive the system into operating based on false information, such as incorrect sensor readings or erroneous control commands. However, differently from arbitrary data injection attacks, special care is taken to ensure the desired degradation effect on the target, often seeking not only accuracy but also attack stealthiness. To be successful, the attack algorithm must not only have knowledge about the system but also read and inject data into the CPS control loops.

The existing intelligent attack techniques aimed to learn information about CPSs and then perform attack actions are detailed in Section \ref{subsec:CPI_Attacks} and \ref{subsec:MB_Attacks}.

\subsection{Cyber-Physical Intelligence Attacks}\label{subsec:CPI_Attacks}

As discussed in Section \ref{subsec:taxonomy}, the CPI class of attacks encompasses three types of attacks: eavesdropping, PSI attacks, and ASI attacks. Eavesdropping corresponds to simply capturing the communication traffic transmitted between the CPS devices. The PSI and ASI attacks, in turn, aim to learn models from the eavesdropped data and are detailed in Sections \ref{subsubsec:PSI_Attacks} and \ref{subsubsec:ASI_Attacks}, respectively.

\subsubsection{Passive System Identification Attacks}\label{subsubsec:PSI_Attacks}

\begin{figure}[ht!]
    \centering
    \includegraphics[trim=1.5cm 6.2cm 2.5cm 4.2cm, clip=true, width=.8\linewidth]{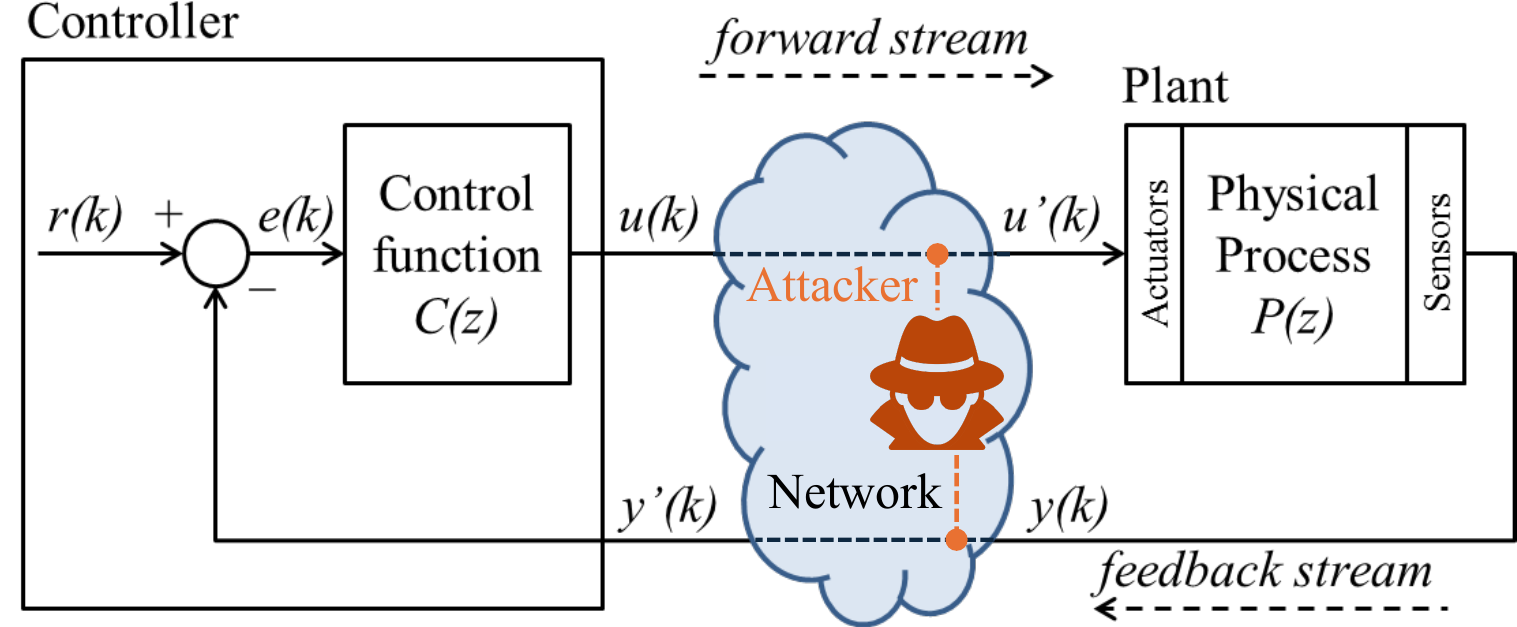}
    \caption{Passive System Identification attack.}
    \label{fig:PSI}
\end{figure}

Passive System Identification (PSI) attacks proposed in \cite{de2017covert} are a form of cyber-physical intelligence gathering that targets CPSs without causing direct interference on the system, thus maintaining a high degree of stealthiness. These attacks, as shown in Figure~\ref{fig:PSI}, involve the passive eavesdropping of communications between controllers, actuators, and sensors within a system to collect information necessary to learn the system's models. The learned models can represent linear and time-invariant transfer functions (or even non-linear functions) that describe the physical behavior of the plant, automation and control algorithms, and network characteristics.

One of the defining features of PSI attacks is their stealthiness. Unlike active attacks, which might introduce observable perturbations into system operations, passive attacks rely solely on the data that naturally flows through the system to learn the desired models. The passive nature of these attacks means they do not actively inject any signals into the system, relying instead on the analysis of ongoing communications to reverse-engineer the system's models. By passively observing the data exchanged between the controller and the plant, attackers can reconstruct a mathematical representation of the system dynamics. This approach minimizes the risk of detection as it does not change the behavior or the performance of the system. This stealthiness is critical as it allows the attack to remain undetected for a longer period, thereby increasing the chance of learning the systems' models without alerting the system operators or triggering security mechanisms.

However, passive system identification attacks come with inherent limitations. This attack, although stealthy, is time-consuming and can be ineffective during periods of steady system states where the transmitted data does not carry information rich enough for the identification process \cite{de2017bio}. 
Their success heavily depends on the quality and quantity of observable data. If the system does not exhibit sufficient variation in its natural operation or if the data is sparse, the accuracy of the identified model can be compromised \cite{de2017covert}. Additionally, these attacks assume that the system's dynamics do not change significantly over time. If the system adapts or if there are any significant alterations in its operation, the model inferred from the attack may quickly become obsolete, and the attack needs to be executed again to learn the target models.

The process of learning the model of the attacked system requires algorithms capable of estimating system parameters from regular operational data.
The algorithms used in passive system identification often include soft computing techniques, which provide robust mechanisms for dealing with both the linear and nonlinear complex nature of most control systems and plants. These algorithms are adept at finding optimal or near-optimal solutions for system identification problems, even when the system exhibits highly complex and stochastic behaviors. The following soft computing algorithms have been employed in passive system identification attacks in order to learn the models of the target system:
\begin{itemize}
    \item Backtracking Search Optimization Algorithm (BSA) \cite{civicioglu2013backtracking}: the BSA is a bio-inspired metaheuristic belonging to the class of evolutionary algorithms. Its bio-inspired approach mimics groups of social animals that sporadically return to previously successful hunting grounds to gather food. The algorithm is designed for solving real-valued numerical optimization problems. BSA differentiates itself by employing a unique memory feature that utilizes a historical population from a previous generation to influence the current search direction, thereby enhancing the algorithm's capability to explore and exploit the solution space effectively. It simplifies parameter control by relying on a single parameter that adjusts the amplitude of the search direction, reducing sensitivity to parameter initialization. The algorithm is structured into several phases, including initialization, mutation, crossover, and selection, which collectively support its robust performance across a variety of benchmark problems. In the context of passive system identification attacks, BSA is utilized to model the system under attack as an optimization problem, where the algorithm adjusts an estimated linear time-invariant (LTI) transfer function is the form of equation (\ref{equ:q_z_tf}): 
    \begin{equation}
    Q(z)=\frac{O(z)}{I(z)} = \frac{a_nz^{n}+a_{n-1}z^{n-1}+...+a_{1}z^{1}+a_{0}}{z^{m}+b_{m-1}z^{m-1}+...+b_{1}z^{1}+b_{0}},
    \label{equ:q_z_tf}
    \end{equation}
    until its output matches the observed output of the actual system. This approach involves capturing input $i(k)=Z^{-1}[I(z)]$ and output $o(k)=Z^{-1}[O(z)]$ data from the system, applying these to the estimated model, and iteratively adjusting the coefficients $[a_n,a_{n-1},...a_1, a_0]$ and $[b_{m-1},b_{m-2},...b_1, b_0]$ of the model to minimize the difference between the model's output and the actual system output.
    \item Particle Swarm Optimization (PSO) \cite{kennedy1995particle}: the PSO is a computational method inspired by social behaviors observed in nature, such as bird flocking and fish schooling. However, the individuals of its population are referred to as particles for the sake of generalization. The fundamental concept of PSO involves a swarm of candidate solutions (particles) moving through the solution space to find the optimal solution. Each particle adjusts its trajectory based on its own experience and the experience of its neighbors or the entire swarm. The movement of each particle is influenced by the best position it has found (personal best) and the best position any particle in the neighborhood has found (global best). This allows PSO to effectively explore and exploit the solution space in search of the optimum solution. In the context of passive system identification attacks, the PSO acts as the BSA. It is used to model the CPS devices by identifying the parameters of their LTI models (such as described in equation (\ref{equ:q_z_tf})) based solely on the system's observable inputs $i(k)=Z^{-1}[I(z)]$ and outputs $o(k)=Z^{-1}[O(z)]$. Note that in both BSA-based and PSO-based system identification attacks, it is necessary to know the order $n$ and $m$ of the LTI transfer function polynomials to define the number of dimensions of the search space explored by the optimization algorithm. Although this is an attack constraint of this kind of approach, this information may be inferred if the attacker, at least, knows what the attacked plant is and what type of its controller. Moreover, this approach presents restrictions to learning models of non-linear targets. 
    \item Artificial Neural Networks (ANN) \cite{graupe2013principles}: ANNs are computational models inspired by the human brain's neural networks, used extensively in pattern recognition and machine learning tasks. They consist of interconnected layers of nodes 8referred to as neurons), where each connection can transmit a signal from one neuron to another. The signal at a connection is modified by a weight, and each neuron can apply a non-linear transformation to its input signals before passing it on to the next layer. The use of ANNs in PSI attacks to learn models model nonlinear plants, such as battery energy storage systems (BESS) \cite{pasetti2021artificial}, offers significant advantages primarily due to their ability to capture and model complex nonlinear relationships and dynamics that traditional linear models cannot. In this way, PSI attacks based on ANN can overcome the restrictions that attacks based on BSA and PSO present, as they are aimed at LTI systems. Note that ANN-based PSI attacks can learn from operational data to predict CPS models without prior explicit programming of the system's dynamics. The drawback is that the models produced by ANN-based attacks are less explainable than those produced by BSA and PSO-based attacks found in the literature whose results translate into expressions like equation (\ref{equ:q_z_tf}).
\end{itemize}

Examples of PSI attacks are found in \cite{de2017covert,ferrari2020model,de2019bio,de2018evaluation} using the BSA to learn the LTI models of CPSs. In \cite{de2017covert,ferrari2020model}, the PSI attack is used to learn the physical model of a DC motor and its control algorithm. In \cite{de2017covert} it is used to support the execution of SD-CDI attacks while in \cite{ferrari2020model} it supports the execution of SD-CDL attacks. In \cite{de2019bio}, the BSA-based PSI attack is implemented together with a noise processing technique to use noisy data on the task of learning generic LTI models. In \cite{de2018evaluation} the PSI attack is used to learn the linearized models of a large pressurized heavy water reactor. The model learned is then used to implement a Man-in-the-Middle (MitM) covert misappropriation attack classified as SD-CDI.

In \cite{pasetti2021artificial,de2022ann} the authors present the use of ANN in PSI attacks to learn the model of a nonlinear plant encompassing physical devices and multi-level controllers with unknown control algorithms. Specifically, the ANN-based PSI attack is used to learn the model of a real BESS, which is coupled with a photovoltaic power plant. Essentially, the attacker intercepts the communication between the BESS and its controller. By employing ANN, the attack code learns and replicates the BESS behavior. This model is then used to support a stealth SD-CDI attack implemented by a MitM.

It is possible to establish an analogy between the PSI attack and the Known Plaintext cryptanalytic attack \cite{stallings2015computer}. As summarised in Table \ref{tab:criptanalise_PSI}, wherein $i(k)$ and $o(k)$ correspond to the known plaintext and cyphertext, respectively, the form of the generic transfer function $Q(z)$ corresponds to the encryption algorithm and the actual coefficients $[a_n,a_{n-1},...a_1, a_0]$ and $[b_{m-1},b_{m-2},...b_1, b_0]$ correspond to the secret key \cite{de2017covert}.

\begin{table}[h]
    \centering
    \begin{tabular}{ll}
    \noalign{\hrule height 1pt}
    Known plaintext attack                & PSI attack \\
    \noalign{\hrule height 1pt}
    Known Plaintext 		 			&	$i(k)=Z^{-1}[I(z)]$\\
    \hline
    Cypher text 					&	$o(k)=Z^{-1}[O(z)]$ \\
    \hline
    Encryption Algorithm 			&	$Q(z)=\frac{O(z)}{I(z)} = \frac{a_nz^{n}+a_{n-1}z^{n-1}+...+a_{1}z^{1}+a_{0}}{z^{m}+b_{m-1}z^{m-1}+...+b_{1}z^{1}+b_{0}}$ \\
    \hline
    Cryptographic key  				&	actual coefficients of $Q(z)$. \\
    \noalign{\hrule height 1pt}
    \end{tabular}
    \caption{Analogy between Known plaintext cryptanalysis attack and the PSI attack}
    \label{tab:criptanalise_PSI}
\end{table}

\subsubsection{Active System Identification Attacks}\label{subsubsec:ASI_Attacks}

The ASI attack was proposed in \cite{de2017bio} to overcome the constraint of the PSI attack in situations where the attacker may not wait so long for the occurrence of meaningful signals in the CPS communication links. Recall that, as discussed in Section \ref{subsubsec:PSI_Attacks}, the PSI attack can be ineffective during periods when the system remains in a steady state, and the eavesdropped data does not carry information rich enough for the identification process. In this sense, the ASI attacks present a distinct advantage over passive approaches by enabling attackers to interact directly with the system, injecting specially crafted signals into it to obtain desired responses. This proactive interaction allows the learning process to occur at the convenience of the attacker, requiring eavesdropping periods not as long as in passive methods which solely rely on observed data without interference. To stimulate the target system and obtain the desired responses, the attacker, acting as an MitM, injects an attack signal $a(k)=Z^{-1}[A(z)]$ (typically an impulse) in the CPS and collects the system's response $y(k)=Z^{-1}[Y(z)]$. Based on these two signals, the attacker then runs the identification process, as in the PSI attack. The learning process can be implemented, for instance, using the same kind of soft computing algorithms described in Section \ref{subsubsec:PSI_Attacks} to learn LTI models in the form of equation (\ref{equ:q_z_tf_ASI}):
\begin{equation}
    Q(z)=\frac{Y(z)}{A(z)} = \frac{a_nz^{n}+a_{n-1}z^{n-1}+...+a_{1}z^{1}+a_{0}}{z^{m}+b_{m-1}z^{m-1}+...+b_{1}z^{1}+b_{0}},
    \label{equ:q_z_tf_ASI}
\end{equation}

\begin{figure}[ht!]
    \centering
    \includegraphics[trim=3cm 5.3cm 4.3cm 4.7cm, clip=true, width=.8\linewidth]{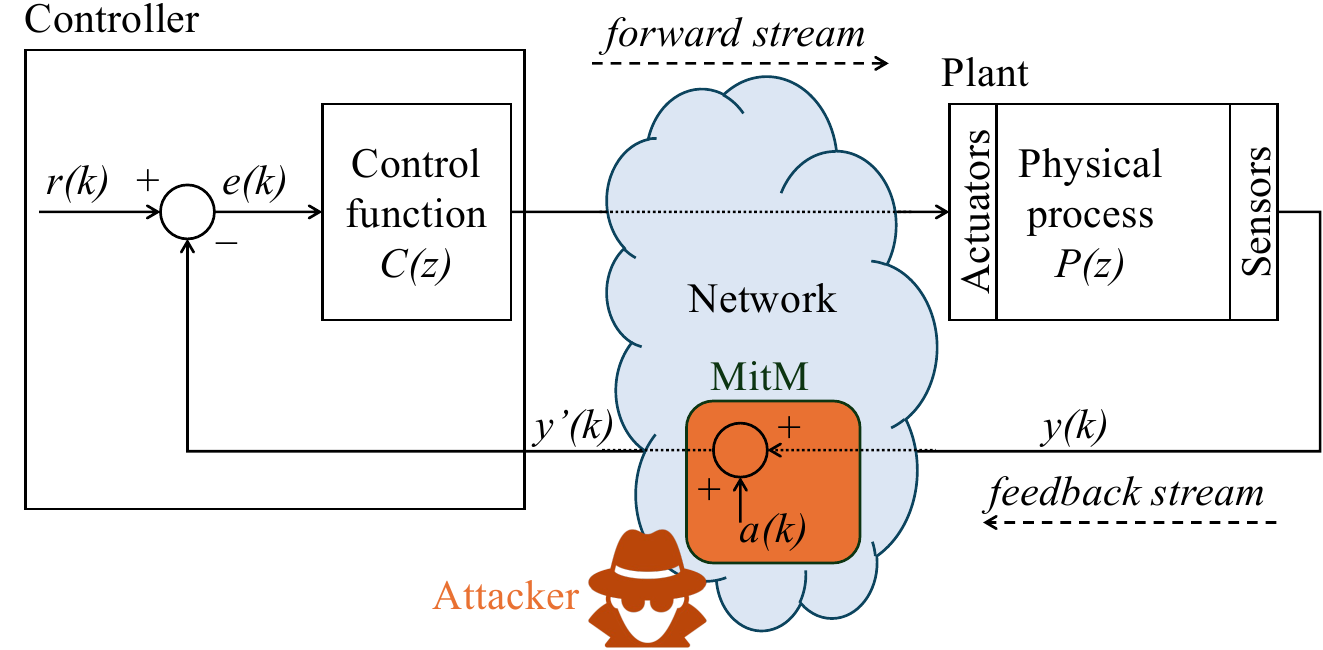}
    \caption{Active System Identification attack.}
    \label{fig:ASI}
\end{figure}

The ASI attacks require a higher level of access to the target system than the PSI attacks, as they need to manipulate signals directly. This can be a challenging prerequisite in high-security environments or systems with limited accessibility. The requirements for carrying out such attacks include the ability to inject data into the system and the capability to measure and analyze the resulting outputs. It means that to inject data, the attacker needs to be able to overcome data integrity checks if they are present in the system. The attacker also needs to have the ability to overcome possible data confidentiality mechanisms in place to analyze the system response signals.

Despite the advantage of not depending on the occurrence of events that are beyond the attacker's control (as required in the PSI attack), the ASI attacks come with a drawback. The visibility of these attacks, due to their interactive nature, makes them susceptible to detection. Not that, unlike passive attacks, ASI attacks are typically less stealthy. By design, these attacks involve injecting signals \cite{de2017bio} or altering system operation parameters \cite{phillips2019covert}, which can be detected by monitoring systems or noticeable performance degradation. This lack of stealth can limit the practical applications of active attacks in environments where security systems are robust and continuously monitored. In summary, active system identification attacks offer the advantage of direct interaction for quicker and more precise system modeling but are hindered by their non-stealthy nature and the high access requirements. While powerful, these attacks are best suited for environments where they can be executed without severe repercussions from detection.

The literature presents examples of ASI attacks \cite{de2017bio,phillips2019covert} that must be highlighted. In the first example of ASI attack \cite{de2017bio}, two algorithms, the BSA and the PSO, were used to learn the models of a CPS composed of a DC motor and a Proportional-Integral (PI) controller. To enhance the accuracy of the identified models, a statistical refinement technique is applied to the results of the metaheuristics. The work demonstrates the attack's effectiveness in supporting the design of other complex attacks such as the SD-CDI.

In \cite{phillips2019covert}, the authors describe an ASI attack on constant setpoint control systems by targeting the proportional-integral-derivative (PID) gain values. This method manipulates the PID derivative parameter to expose the dynamic characteristics that are typically subdued in constant setpoint operations. By doing so, the attack enhances the accuracy of the system identification process, which is crucial for the accurate modeling of the system's dynamics. The attack is designed to be covert, aiming to evade detection by physics-based anomaly detection systems, which typically look for deviations from expected physical behaviors. The approach leverages brief, calculated adjustments to the PID settings. Its effectiveness and stealth are validated through simulations in Simulink, considering an inverted pendulum system as the target. However, stealth validation of the attack proposed in \cite{phillips2019covert} requires a more comprehensive analysis in terms of state-of-the-art anomaly detection algorithms.

In contrast to the PSI attack (which is understood as analogous to the Known Plaintext cryptanalytic attack), the ASI attack is understood as analogous to the Chosen Plaintext cryptanalytic attack \cite{stallings2015computer}. As summarised in Table \ref{tab:criptanalise_ASI}, wherein $a(k)$ and $y(k)$ correspond to the chosen plaintext and cyphertext, respectively, the form of the generic transfer function $Q(z)$ corresponds to the encryption algorithm and the actual coefficients $[a_n,a_{n-1},...a_1, a_0]$ and $[b_{m-1},b_{m-2},...b_1, b_0]$ correspond to the secret key \cite{de2017covert}.

\begin{table}[h]
    \centering
    \begin{tabular}{ll}
    \noalign{\hrule height 1pt}
    Chosen plaintext attack                & ASI attack \\
    \noalign{\hrule height 1pt}
    Chosen plaintext 			&	$a(k)=Z^{-1}[A(z)]$\\
    Cypher text 					&	$y(k)=Z^{-1}[Y(z)]$ \\
    Encryption Algorithm 			&	$Q(z)=\frac{Y(z)}{A(z)} = \frac{a_nz^{n}+a_{n-1}z^{n-1}+...+a_{1}z^{1}+a_{0}}{z^{m}+b_{m-1}z^{m-1}+...+b_{1}z^{1}+b_{0}}$ \\    
    Cryptographic key  				&	actual coefficients of $Q(z)$. \\
    \noalign{\hrule height 1pt}
    \end{tabular}
    \caption{Analogy between chosen plaintext cryptanalysis attack and the ASI attack}
    \label{tab:criptanalise_ASI}
\end{table}

\subsection{Model-based Attacks}\label{subsec:MB_Attacks}

This section presents the existing model-based attacks that rely on the knowledge obtained by system identification attacks to achieve harmful abilities, such as accuracy, stealthiness, and some degree of decision autonomy.

\subsubsection{SD-CDI Attacks}\label{subsubsec:SD-CDI_Attacks}

According to the classification detailed in Section \ref{subsec:taxonomy}, the attack discussed in this section is referred to as an SD-Controlled Data Injection attack \cite{de2017covert}. This model-based attack aims to decrease the MTBF of the plant and/or impair the efficiency of the plant's physical processes by introducing false data into the control loop of a CPS.

\begin{figure}[h!]
    \centering
    \includegraphics[trim=3cm 5.5cm 4.1cm 4.7cm, clip=true, width=.75\linewidth]{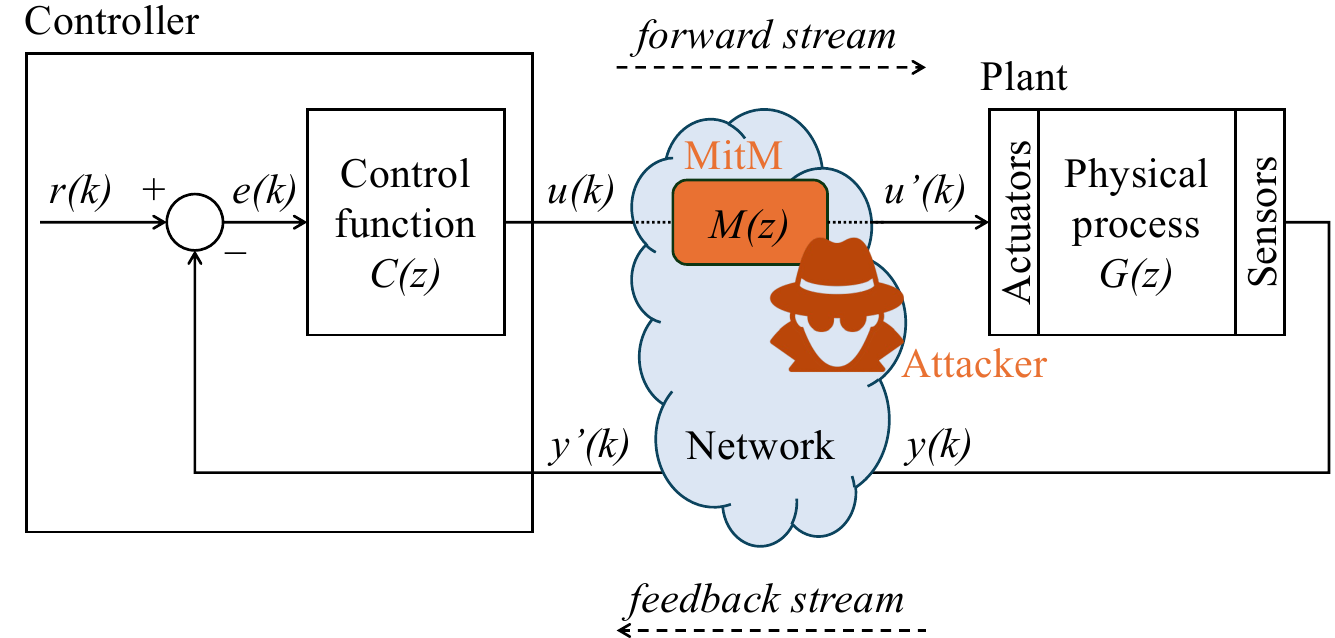}
    \caption{SD-CDI attack.}
    \label{fig:SD-CDI_simple}
\end{figure}

A practical method to undermine a plant's physical operation includes inducing an overshoot during its transient response. These overshoots can lead to stress and potential damage in physical systems, such as mechanical, chemical, and electromechanical systems \cite{el1989variable,tran2007robust}. Often quick and transient, these overshoots are typically hard for humans to detect. Another tactic involves generating a persistent steady-state error in the plant, which can subtly compromise process efficiency or cause long-term damage \cite{de2017covert}.

In this attack technique, to cause either an overshoot or a steady-state error, the attacker manipulates the CPS communication links by strategically injecting false data. This is accomplished through a Man-in-the-Middle (MitM) attack where the attacker executes a function $M(z)$, depicted in Figure~\ref{fig:SD-CDI_simple}, modifying for instance the control signals $u(k)$ to $u'(k)$. It's important to note from Figure \ref{fig:SD-CDI_simple} that while the MitM is positioned in the forward stream, interference could also occur within the CPS feedback stream.

This SD-CDI attack is part of a dual-stage operation alongside a System Identification attack:
\begin{enumerate}[leftmargin=.9in]
	\item [STAGE-I:] The attacker conducts a PSI or ASI attack, as detailed in Section \ref{subsec:CPI_Attacks}, to accurately learn the transfer function $G(z)$ of the plant and the control function $C(z)$ of the controller.
	\item [STAGE-II:] With this knowledge, the attack code, posing as a MitM, introduces manipulated data into the CPS control loop based on the attack function $M(z)$ to subtly alter the plant's physical operations in an accurate and controlled way.
\end{enumerate}

The success of the attack heavily relies on the design of $M(z)$, which depends on the accuracy of the model learned through the system identification attack in STAGE-I. Examples of SD-CDI attacks presented in the literature include \cite{de2017covert,de2017bio}. In \cite{de2017covert} the SD-CDI uses the models learned through a PSI attack to cause overshoots and steady-state errors in a DC motor, while in \cite{de2017bio} the SD-CDI attack is designed based on models learned through an ASI attack.

\subsubsection{SD-CDI Attacks with covert misappropriation architecture}\label{subsubsec:SD-CDI_Covert_Attacks}

\begin{figure}[h!]
    \centering
    \includegraphics[trim=3cm 6.4cm 4.3cm 4.2cm, clip=true, width=.8\linewidth]{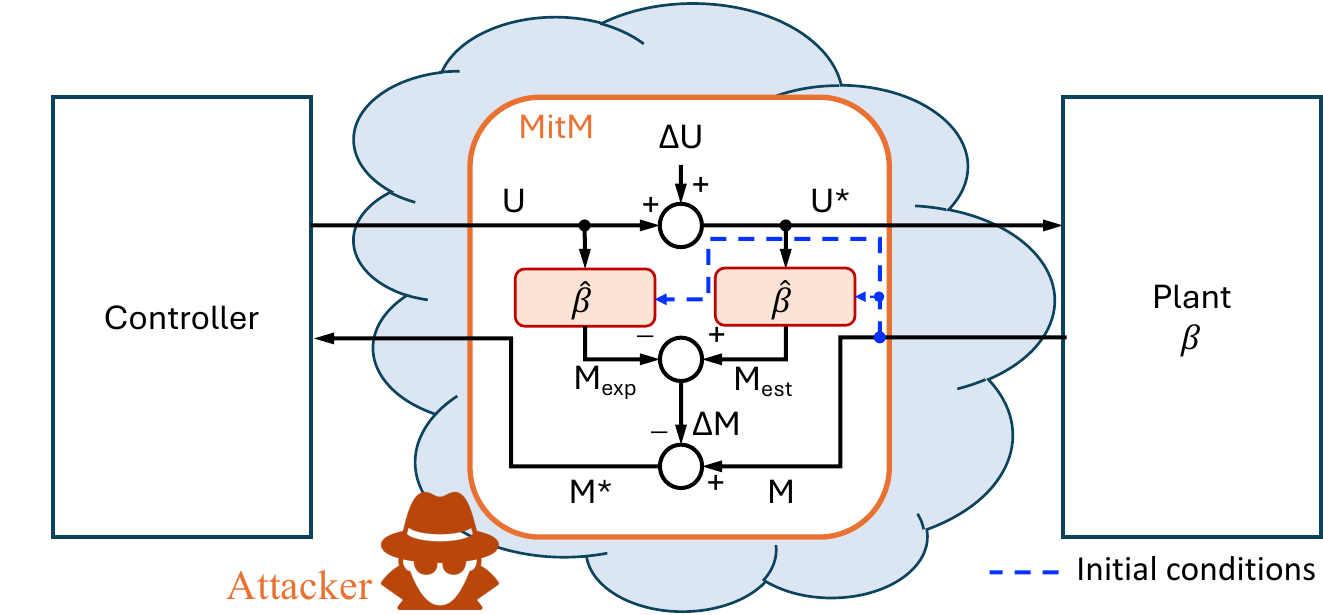}
    \caption{SD-CDI attack with covert misappropriation architecture.}
    \label{fig:SD-CDI}
\end{figure}

The concept of a covert misappropriation attack targeting Networked Control Systems (NCSs) was first introduced in \cite{smith2015covert}. This type of attack enables a Man-in-the-Middle (MitM) to execute unauthorized control operations on a physical system without being detected by the system's legitimate controller. An example of this attack is illustrated in Figure \ref{fig:SD-CDI}. In this attack architecture, $U^*$ denotes the manipulated control signal sent by the attacker to affect the plant behavior, $M^*$ represents the fake feedback signal sent by the attacker to the controller to hide the effects caused by the attack on the plant, and $\hat{\beta}$ is the model learned by a system identification attack to make the data injection stealth.

In the diagram, it's noted that the MitM manipulates the input to the plant as shown in equation (\ref{equ:U}):
\begin{equation}
U^* = U + \Delta U,
\label{equ:U}
\end{equation}
wherein $U$ is the control signal sent by the legitimate controller, and $\Delta U$ is the disturbing signal defined according to the attacker's will. The feedback is manipulated according to equation (\ref{equ:M}):
\begin{equation}
M^* = M - \Delta M,
\label{equ:M}
\end{equation}
wherein $M$ is the legitimate feedback signal, and $\Delta M = M_{est} -  M_{exp}$ represents the estimated disturbance caused by the attack on the plant. Note that $M_{est}$ is the estimation of the plant output $M$ as a result of $U^*$, and $M_{exp}$ is what would be expected as plant output $M$ if $U$ was applied to the plant without attack. The accuracy of $M_{est}$ and $M_{exp}$ and, thus, the stealthiness of the attack depends on the accuracy of $\hat{\beta}$. This MitM architecture ensures that, if the attacker accurately knows the plant model ({\it i.e.} $\hat{\beta} \longrightarrow \beta$), then the altered feedback mimics normal operation signals ({\it i.e.} $M^* \longrightarrow M$), deceiving the controller into thinking no attack is occurring.

Examples of SD-CDI Attacks with covert misappropriation architecture in the literature include \cite{smith2015covert,de2018evaluation,pasetti2021artificial,de2022ann}. In \cite{smith2015covert} the attack is shown in linear and nonlinear plants such as an irrigation canal and a DC motor, respectively. Despite proposing the concept of covert misappropriation attack, work \cite{smith2015covert} does not include mechanisms for learning the model required to achieve stealth. This is achieved in \cite{de2018evaluation}, where an intelligent attack is composed of a model learning step using a BSA-based PSI attack and a covert misappropriation architecture. The BSA-based PSI attack is used to learn the linearized model of a large pressurized heavy water reactor, which is then embedded in the covert MitM architecture. 

Another example of a covert SD-CDI Attack is presented in \cite{pasetti2021artificial,de2022ann}. The paper demonstrates that traditional system identification approaches are typically suited only for linear time-invariant systems and are not effective for nonlinear systems like those found in battery energy storage setups (such as a BESS). By implementing ANNs, the system identification can handle the nonlinear characteristics inherent in such systems, which involve various physical devices and control levels with unknown control algorithms. In this case, the covert architecture embeds the trained ANN models to mimic the behavior of a real BESS while harmful behaviors are caused on the system. The MitM implementation of \cite{pasetti2021artificial} is shown feasible in Modbus/TCP networks.

\subsubsection{SD-CDL Attacks}\label{subsubsec:SD-CDL_Attacks}

\begin{figure}[h!]
    \centering
    \includegraphics[trim=3cm 7.1cm 4.3cm 4.7cm, clip=true, width=.75\linewidth]{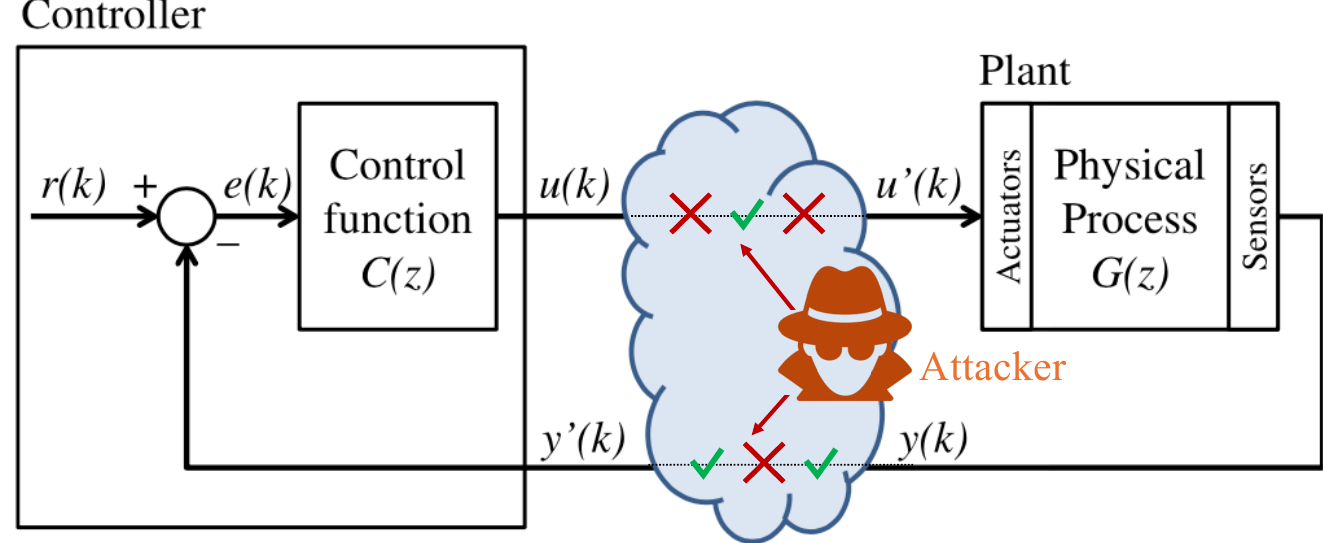}
    \caption{SD-CDL attack.}
    \label{fig:SD-CDL}
\end{figure}

The SD-CDL attack discussed in this section was firstly described in \cite{ferrari2020model} as an intelligent attack method to degrade the service of CPSs based on Real-Time Ethernet (RTE) protocols. To affect the plant behavior, the attack shown in Figure \ref{fig:SD-CDL} causes selective data losses in the CPS communication network, carefully avoiding massive data loss or complete communication interruption to remain stealthy.

A feasible method to compromise plant operations using SD-CDL attacks is to induce overshoots, potentially stressing or damaging the system over time. Alternative behaviors, such as extended settling times, are also achievable through different data loss strategies.

The SD-CDL attack involves selectively dropping packets that contain crucial samples in both forward and feedback streams of the CPS through tactics like packet injection in PROFINET networks \cite{ferrari2020model}. To intelligently decide which packets to drop to cause a specific malicious behavior, the attack code uses the BSA metaheuristic and the CPS models previously learned through a system identification attack. The attack sequence ({\it i.e.}, the sequence of packets to drop) is defined with the BSA by minimizing a fitness function that ensures the system's output approaches the desired effect.

Examples of SD-CDL attacks in the literature include \cite{ferrari2020model,de2021soft}. In \cite{ferrari2020model} the authors show the feasibility of this intelligent attack to cause overshoots on a DC motor. The attack was executed in PROFINET network implemented with commercially available industrial hardware (Siemens SIMATIC S7-1500 PLCs and a Siemens Scalance XB208 industrial switch). In \cite{de2021soft} the authors compare the performance of the BSA and the PSO algorithms to find effective attack sequences. They also redesign the fitness function optimized by these metaheuristics to not only produce accurate attacks but also to reduce even more the number of packets to be dropped (thus, increasing the attack stealthiness). These works show that algorithms like BSA and PSO can be used not only to learn the models of the targeted CPSs but also to autonomously decide effective and optimized attack sequences in the field.

\begin{table}[h!]
    \centering
    \begin{tabular}{p{0.01\textwidth}p{0.03\textwidth}p{0.02\textwidth}ccccc}
        \noalign{\hrule height 1pt}
             & &  \multicolumn{2}{c}{Learning Strategy}  &  \multicolumn{2}{c}{Model-Based Attack}  &  & \\
        \cline{3-4} \cline{5-6}
        Ref. & Year & Type & Algorithm & Type & Algorithm & Target & Protocol \\
        \noalign{\hrule height 1pt}
        \cite{de2017covert} & 2017 & PSI & BSA & SD-CDI & LTI Transfer & DC motor & -- \\
                     &  & &  &  &  Function &  &  \\
        \hline
        \cite{de2017bio} & 2017 & ASI & PSO / BSA & SD-CDI & LTI Transfer & DC motor & -- \\
                             &  & &  &  &  Function &  &  \\
        \hline
        \cite{de2018evaluation}     & 2018  & PSI & BSA & SD-CDI & Covert & Pressurized Heavy & -- \\
             &  & &  &  &  Misappropriation & Water Reactor &  \\
        \hline
        \cite{de2019bio}     & 2019 & PSI & BSA with  & -- & -- & Generic & -- \\
            &  & & noise &  &  &  &  \\
            &  & & processing &  &  &  &  \\
        \hline
        \cite{phillips2019covert}     & 2019 & ASI & PID with & -- & -- & Inverted & -- \\
                    &  & & reduced &  &  & pendulum &  \\
                    &  & & derivative gain &  &  &  &  \\
        \hline
        \cite{ferrari2020model} & 2020 & PSI & BSA & SD-CDL & BSA & DC motor & PROFINET \\
        \hline
        \cite{de2021soft}   & 2021 & -- & -- & SD-CDL & PSO / BSA & DC motor  & -- \\
        \hline
        \cite{pasetti2021artificial} & 2021 & PSI & Feedforward  & SD-CDI & Covert           & Battery Energy & Modbus TCP\\
                                     &      &     & ANNs         &        & Misappropriation & Storage System &  \\
        \hline                             
        \cite{de2022ann}             & 2022 & PSI &  Feedforward & SD-CDI & Covert           & Battery Energy &  -- \\
                                     &      &     & ANNs         &        & Misappropriation & Storage System &  \\
        \noalign{\hrule height 1pt}
    \end{tabular}
    \caption{Summary of the discussed attacks in CPSs.}
    \label{tab:summary_attacks_CPS}
\end{table}

\section{Intelligent Attacks in IT Environments}\label{sec:IT}

Since most OT cyberattacks actually start in IT networks before pivoting into OT, it is worth studying the possible use of AI techniques to develop intelligent and automated attacks in IT environments. 
In the field of IT networks, recent research shows the use of Artificial Intelligence (AI) to develop machine learning (ML)-based penetration testing (PT) automation tools. While these tools are designed to enhance the security of networked systems, they can also rapidly become attack tools accessible to potential attackers. 

In \cite{Valea_2020}, the authors propose a PT tool based on a decision tree (DT) algorithm to autonomously perform network port scanning, search for software vulnerabilities, and select suitable exploitation methods. 
The DT algorithm, which has been trained using a dataset that includes parameters such as service, port, CVEs, exploit, and operating system, predicts the most appropriate exploit for the targets.  
Experimental findings indicate that the model achieves a 33\% accuracy rate in predicting exploits. Although this performance is not considered high, it does imply some capability in automating standard penetration testing procedures, considering that the DT must make predictions across multiple classes. 

In \cite{Jagamogan_2022}, the study evaluates GyoiThon, a penetration testing tool, by examining its standard and machine learning (ML) modes, specifically the Na\"{\i}ve Bayes approach. The research involved testing websites both with and without a Content Management System (CMS). Results indicate that GyoiThon's ML mode surpasses the standard mode in detecting a wider array of vulnerabilities. 
The study supports the hypothesis that incorporating ML algorithms improves PT tools.

Note that both approaches \cite{Valea_2020,Jagamogan_2022} use simple ML solutions, such as DT and Na\"{\i}ve Bayes. Although the results are satisfactory, the solutions have problems related to the low amount of data used for training and for evaluating these models. The bias-variance trade-off or overfitting was reported by the authors as problems related to a few data for modeling. However, nowadays, it is reasonable to assume that there are ways to collect more data in order to use Deep Learning-based solutions, which have the potential to leverage better results.

In \cite{Ghanem_2020}, the research describes an AI-driven automated PT system that uses reinforcement learning (RL) to improve PT performance. This system is designed around a Partially Observable Markov Decision Process (POMDP), incorporating essential environmental data and PT tasks. The findings demonstrate that this approach successfully mimics human expertise and performs PT tasks autonomously or with minimal human input, achieving better results than traditional manual PT in small network environments. 
In \cite{Hu_2020}, the authors propose an automated PT framework based on deep reinforcement learning, specifically employing a Deep Q-Learning Network (DQN). This framework utilizes the Shodan search engine and the MulVAL algorithm to gather actual host and vulnerability data, thereby creating a realistic dataset for DQN model training. The DQN model is trained to identify the best attack pathways, achieving an 86\% accuracy rate in test scenarios. The findings highlight the framework's capability to propose attack strategies and its potential to be integrated into real-world PT tools.

The aforementioned studies highlight not only the potential for utilizing state-of-the-art PT automation tools in intelligent attacks but also indicate a trend toward these tools consistently improving their performance. Note that these AI-based technologies may have the potential not only to facilitate attacks on critical infrastructures that adopt IT but they can also pave the way for the deployment of intelligent attack codes in OT environments.

\begin{table}
    \centering
    \begin{tabular}{ccccc}
    \noalign{\hrule height 1pt}
        Ref. & Year & Algorithm & Tools & Dataset coverage \\
    \noalign{\hrule height 1pt}
       \cite{Valea_2020}  & 2020 & Decision Tree & Metasploit & 160 devices\\
    \hline
       \cite{Jagamogan_2022}  & 2022 & Na\"{\i}ve Bayes & GyoiThon  & --- \\
    \hline
       \cite{Ghanem_2020}  & 2020 & Reinforcement Learning & --- & 100 devices\\
    \hline
       \cite{Hu_2020}  & 2020 & Reinforcement Learning (DQN) & Shodan &  2000 different attack trees\\
    \noalign{\hrule height 1pt}
    \end{tabular}
    \caption{Summary of the discussed intelligent attacks IT environments.}
    \label{tab:summary_attacks_IT}
\end{table}

\section{Discussion}\label{sec:disc}

An owner of a CPS may assume their system is protected against intelligent attacks, believing that potential attackers lack knowledge of the system’s design and its models. However, the attacks on CPSs studied in this chapter, and summarised in Table \ref{tab:summary_attacks_CPS}, reveal that these threats can be initiated with limited information about the target. These works point to the emergence of attacks capable of learning information and models from the target system and then using this knowledge to act accurately, stealthily, and even autonomously. Therefore, it is critical that system security measures remain stringent and that protective actions are implemented.

As depicted in Figure \ref{fig:attack_roadmap}, an intelligent model-based attack in a CPS unfolds through a series of three distinct phases: eavesdropping, system identification (either active or passive), and model-based disruption (such as controlled data injection, packet loss, or jitter). The requirements specified in Figure \ref{fig:attack_roadmap} aid in formulating multi-layered defense strategies against these intelligent attacks. Consequently, a comprehensive set of preventative actions can be designed based on these specifications:
\begin{enumerate}
\item [I -] An initial and straightforward mitigation strategy is to prevent unauthorized access to the CPS communication network, which, as shown in Figure \ref{fig:attack_roadmap}, could thwart all three attack stages. Segregating the OT network from other networks is considered an effective architectural safeguard for a CPS, as recommended in \cite{stouffer2015nist}. However, when network segregation isn’t practical or desired, reducing potential unauthorized access might be achieved through network segmentation, demilitarized zones (DMZ), firewall policies, and adopting specific network designs as outlined in \cite{stouffer2015nist}.
\item [II -] Beyond securing the control loop, additional safeguards should prevent unauthorized data access if initial measures fail. In \cite{pang2012design}, a secure transmission mechanism combining symmetric-key encryption, hash algorithms, and timestamp techniques is proposed to ensure data confidentiality, integrity, and authenticity between the CPS devices. Implementing such measures would obstruct access to the CPS data, necessary for executing system identification and intelligent model-based data injections.
\item [III -] Preventing an intelligent model-based attack further involves obstructing the attacker’s ability to gain required system knowledge. If an attacker still manages to access the CPS communication links and data, it is essential to make the system identification process more challenging and less accurate. Therefore, the third preventative measure involves hindering the learning process. In this direction, the literature brings solutions where switching controllers \cite{de2017use,de2018controller} are used to affect the performance of system identification attacks. However, mechanisms that hinder the learning process of intelligent attacks are a topic that is still little researched.
\end{enumerate}

It is also important to highlight that the protection of CPS and critical infrastructures involves not only the security of OT networks but also IT networks. In relation to the latter, Section \ref{sec:IT} also points to the emergence of PT automation techniques (summarized in Table \ref{tab:summary_attacks_IT}) with the potential to automate intelligent cyberattacks, and these tools present the tendency to have increasingly better performance. Research into mechanisms that hinder the learning process of this type of tool is still little studied and represents a relevant field to be explored.

\section{Conclusion}\label{sec:concl}

This chapter discusses existing techniques with the potential to be used in the development of intelligent attacks in CPSs and critical infrastructures built on OT and IT. Techniques for intelligent attacks in OT environments include AI-based algorithms to learn models of the targeted systems to conduct stealthy, optimized, and automated attacks. These attacks can degrade the performance or cause failures in a wide range of systems, such as industrial devices, energy systems, and nuclear plants, among others. Moreover, these attacks are proven to be feasible in OT networks built with currently adopted industrial protocols, such as PROFINET and Modbus/TCP. 

Conversely, IT environments witness the use of emerging intelligent PT tools, which, while designed to bolster security, can also be repurposed by malicious actors to facilitate attacks. These tools use ML algorithms to automate and optimize the process of finding and exploiting vulnerabilities, which underscores the dual-use nature of such technologies and the possible threats they pose.

The landscape shown in this chapter encourages future research for new methodologies to hinder the learning process of intelligent attacks, potentially limiting their effectiveness. This can include the use of deceptive techniques to mislead the learning algorithms of the attackers, or the design of systems that can adapt in real-time to prevent intelligent attacks from developing a clear understanding of the system's operations. 

\section{Acknowledgement} \label{sec:ack}


This work was supported by the 'Companhia Paranaense de Energia - COPEL' research and technological development program, through the PD-02866-0533/2021 project, regulated by ANEEL, and by FCT through the LASIGE Research Unit, ref.\ UIDB/00408/2020 (https://doi.org/ 10.54499/UIDB/00408/2020) and ref.\ UIDP/00408/2020 (https://doi.org/10.54499/UIDP/00408/2020).





\bibliographystyle{ieeetr}
\bibliography{biblio}

\end{document}